\documentclass[letterpaper]{article}
\usepackage{fullpage}
\usepackage[dvips]{graphicx}
\usepackage[sumlimits]{amsmath}
\usepackage{graphicx}
\usepackage{multirow}
\usepackage{subfigure}
\usepackage{url}
\usepackage{epsfig}

\begin{document}

\pagestyle{empty}


\title{Approximating Document Frequency with Term Count Values}

\author{Martin Klein, Michael L. Nelson\\
Old Dominion University, Department of Computer Science\\
Norfolk, VA 23529 USA}


\maketitle

\begin{center}
\{mklein, mln\}@cs.odu.edu
\end{center}

\begin{abstract}
%
For bounded datasets such as the TREC Web Track (WT10g) the computation of term frequency (TF) and inverse document frequency (IDF) 
is not difficult.
However, when the corpus is the entire web, direct IDF calculation is impossible and values must instead be estimated.
Most available datasets provide values for \textit{term count (TC)} meaning the number of times a certain term occurs in the entire
corpus. Intuitively this value is different from \textit{document frequency (DF)}, the number of documents (e.g., web pages) a
certain term occurs in.
We conduct a comparison study between $TC$ and $DF$ values within the Web as Corpus (WaC).
We found a very strong correlation with Spearman's $\rho\ge0.8$ ($p\le0.005$) which makes us confident in claiming that for such
recently created corpora the $TC$ and $DF$ values can be used interchangeably to compute IDF values.
These results are useful for the generation of accurate lexical signatures based on the TF-IDF scheme.
\end{abstract}
\section{Introduction and Motivation} \label{sec:intro}
In information retrieval (IR) research the term frequency (TF) - inverse document frequency (IDF) concept is well known and established to
extract the most significant terms while dismissing the more common terms from textual content. It also has been used to generate 
lexical signatures (LSs) of web pages \cite{phelps:hyperlinks,park:ls-tois,harrison:just-in-time,klein:ls,sugiyama:refinement-of-tfidf}.
Such LSs can be used to (re-)discover missing web pages when fed back into search engine interfaces.
The computation of TF values for a web page is straight forward since we can simply count the occurrences for each term within the page.
Two values are mandatory for the IDF computation: the overall amount of documents in the corpus and the amount of documents a term appears in.
We call the second value \textit{document frequency (DF)}.
Since both values are unknown when the entire web is the corpus, accurate IDF computation for web pages is impossible and values need to be
estimated. 

Various corpora containing web pages, their textual content and their in- and outlinks are available and can be used to estimate IDF values since
they are considered a representative sample for the Internet \cite{soboroff:trec}. The TREC Web Track is probably the most common corpus
and has, for example, been used in \cite{sugiyama:refinement-of-tfidf} for IDF estimation. The British National Corpus (BNC) \cite{leech:frequencies},
as another example, has been used in \cite{staddon:inference}.
Google published the N-grams \cite{google-n-grams} in $2006$ and hence provides a powerful alternative source for $TC$ values of terms extracted
from web pages from the Google index.
The \textit{Web as Corpus kool ynitiative (WaCky)} provides the WaC corpus as an alternative with no charge for researchers.
The problem with these corpora is that they do not provide $DF$ values for the terms (or $n$-term tokens) they contain.
We can count the total number of documents and therefore determine the $DF$ values in case the corpus documents are provided along with the terms.
\begin{table}
 \centering
  \begin{minipage}{\textwidth}
 \centering
  \begin{tabular}{|c||c|c|c|c|} \hline
        Corpus&\textbf{Google N-gram}&\textbf{TREC WT10g}&\textbf{BNC}&\textbf{WaC} \\ \hline \hline
        \multirow{6}{*}{}&&&& \\ 
        Source&Google indexed&English language&British English Texts&uk.Domain \\ 
        &English language&Web Pages&(newspapers/journals,&Web Pages  \\
        &Web Pages&&books), Transcripts of& \\
        &&&Verbal Language& \\ 
        &&&(meetings, radio shows)& \\ \hline
        Date&$2006$&$1997$&$1994$&$2006$\\ \hline
        \multirow{3}{*}{}&&&& \\
        Unique&$>13M$&$5.8M$\cite{kolcz:improved_robustness}&N/A&$>10M$ \\ 
        Terms&&&$>100M$ Total Terms&\\ \hline
        \multirow{3}{*}{}&&&& \\
        Number of&$>1B$&$1.6M$&$4,124$&$>2.6M$ \\
        Documents&(Not Available)&(Available)&N/A&(Available) \\ \hline
        \multirow{3}{*}{}&&&&\\ 
        $TC$&Available&Not Available&Available&Available \\ 
        &&&from $3^{rd}$ Party& \\ \hline
        \multirow{3}{*}{}&&&& \\
        Freely&No\footnote{A limited number of free copies of the corpus are available from the Linguistic Data Consortium, University of Pennsylvania}&No&No&Yes\\
	Available&&&& \\ \hline
  \end{tabular}
  \end{minipage}
  \caption{Available Text Corpora Characteristics}
 \label{tab:corpora}
\end{table}
Table \ref{tab:corpora} gives an overview of selected corpora and their characteristics. The first row indicates what kind of documents
the corpus is based upon. The row \textit{Number of Documents} shows the total number of documents the corpus consists of (or in the case
of the Google N-grams the number of documents the corpus was generated from). This row also gives information about whether the documents of
the corpus are available. As mentioned above, recognizing the document boundaries within the corpus becomes necessary when computing IDF values.

The row \textit{TC} indicates whether $TC$ values of the corpus are available. 
The following simple example is to illustrate the difference between $TC$ and $DF$. Let us consider a corpus of $5$ documents $D={d_1...d_5}$
where each document contains the title of a song by The Beatles: \\ \\
$
d_1={Please~Please~Me} \\
d_2={Can't~Buy~Me~Love} \\
d_3={All~You~Need~Is~Love} \\
d_4={All~My~Loving} \\
d_5={Long,~Long,~Long} \\ \\
$
Table \ref{tab:tc_df_example} shows the $TC$ and $DF$ values of all terms occurring in our small sample corpus. We can see that the values
are identical for the majority of the terms ($8$ out of $10$). The example also shows that term processing such as stemming would have an
impact on these numbers since \textit{Love} and \textit{Loving} are here treated as different terms.
\begin{table}
 \centering
  \begin{tabular}{|c||c|c|c|c|c|c|c|c|c|c|c|c|} \hline
        \textbf{Term}&All&Buy&Can't&Is&Love&Me&Need&Please&You&My&Loving&Long \\ \hline
	\textbf{TC}&2&1&1&1&2&2&1&2&1&1&1&3 \\ \hline
	\textbf{DF}&2&1&1&1&2&2&1&1&1&1&1&1 \\ \hline
  \end{tabular}
  \caption{$TC$-$DF$ Comparison Example}
 \label{tab:tc_df_example}
\end{table}

Since we are interested in computing accurate IDF values for web page content it seems reasonable to chose a corpus that is based on textual
content of web pages. Even though the TREC WT10g provides the documents and the corpus size seems sufficiently large, it has been shown to be
somewhat dated \cite{chiang:wt10g}.

We are interested in using the Google N-gram dataset as a corpus to generate accurate IDF values from but unfortunately Google only
provides $TC$ values. The WaC corpus in contrast provides both, $TC$ and $DF$ values and therefore we can:
\begin{enumerate}
\item establish a relationship between $TC$ and $DF$ values within the WaC \label{tc-df}
\item establish a relationship between WaC based $TC$ and Google N-gram based $TC$ \label{tc-tc}
\item and finally infer Google N-gram $DF$ from point \ref{tc-df} and point \ref{tc-tc}.
\end{enumerate}

This paper presents the preliminary results of the study and the results indicate that for sufficiently sized and recently generated corpora
$DF$ values can be estimated from $TC$ values.
\section{Related Work} \label{sec:relwork}
\subsection{Correlation between $DF$ and $TC$ Values}
Zhu et al. \cite{zhu:improving_trigram} used an Internet search engine to obtain estimates for $DF$ values of n-grams. They used these values to estimate
$TC$ values and compared those to $TC$ values from a $103$ million word Broadcast News corpus which acted as their baseline. They found that the values
are very similar and thus conclude that values obtained from the web are usable to estimate $TC$.
Keller et al. \cite{keller:using_the_web} also used Internet search engines to obtain $DF$ values for bigrams. Like Zhu et al. they show a high correlation
between values obtained from the web and values from a given (traditional) corpus (the BNC). The main application Keller et al. suggests is for bigrams that
are missing in a given corpus.
Nakov et al. \cite{nakov:using_se_page_hits} show that the n-gram count from several Internet search engines differs but is not statistically significantly
different. They also show that the results from one search engine are stable over time which is encouraging for researchers using this technique to obtain
$TC$ values.

All these studies have two things in common: 1) they all show a strong correlation between $DF$ and $TC$ values and 2) they use $DF$ estimates from
search engines and compare it to $TC$ values from conventional corpora. This is where our approach is different since we use $TC$ values from 
well established text corpora and show the correlation to measured $DF$ values.
\subsection{Generating IDF Values for Web Pages}
Sugiyama et al. \cite{sugiyama:refinement-of-tfidf} use the TREC-9 Web Track dataset \cite{hawking:trec} to estimate IDF values
for web pages. The novel part of their work was to also include the content of hyperlinked neighboring pages in the TF-IDF calculation
of a centroid page.
They show that augmenting the generation of TF-IDF values with content of in-linked pages increases the retrieval accuracy more than
augmenting TF-IDF values with content from out-linked pages.
They claim that this method represents the web page's content more accurately and hence improves the retrieval performance.

Phelps and Wilensky \cite{phelps:hyperlinks} proposed using the TF-IDF model to generate LSs of web pages and introduced
``robust hyperlinks'', an URL with a LS appended.
Phelps and Wilensky conjectured if the an URL would return a HTTP $404$ error, the web browser could submit the appended LS to a
search engine to either find a copy of the page at a different URL or a page with similar content compared to the missing page.
Phelps and Wilensky did not publish details about how they determined IDF values but stated that the mandatory figures
can be taken from Internet search engines. That implies the assumption that the index of a search engine is representative for
all Internet resources. However, they do not publish the value they used for the estimated total number of documents on the
Internet.
\section{Experiment Setup} \label{sec:exp_setup}
The WaC corpus provides what they call a frequency list, a list of all unique terms in the corpus (lemmatized and non-lemmatized)
and their $TC$ value. Since the document boundaries in the corpus are obvious, we can compute the $DF$ values for all terms.
Since we are interested in generating TF-IDF values for web pages and feeding them back into search engines we dismiss all lemmatized
terms and only use the non-lemmatized terms.
We rank both lists in decreasing order of their $TC$ and $DF$ values and honor ties with the minimum value. 
For example consider four terms \textit{a, b, c, d} where term \textit{a} has the highest value, terms \textit{b} and \textit{c} have
the same second highest value and term \textit{d} has the lowest value. The ranking here would be \textit{a=1, b=2, c=2, d=4}.
This kind of ranking is also known as the \textit{sports ranking}.
We compute the Spearman's $\rho$ and Kendall $\tau$ to mathematically prove the correlation.
Table \ref{tab:tc_df_20} shows the top $20$ terms from the WaC corpus ordered by decreasing $TC$ and $DF$ values. The similarity between
the two rankings already becomes visible with that small example since the intersection of both top $20$ rankings just holds $22$ unique
terms. It is not surprising that the $TC$ values are much greater than the $DF$ values since for $DF$ duplicates within one document are
not counted. 
\begin{table}
 \centering
  \caption{Top $20$ Terms and their $TC$ and $DF$ Values}
  \begin{tabular}{|c||c|c||c|c|} \hline
        \textbf{Rank}&\textbf{Term}&\textbf{$TC$}&\textbf{Term}&\textbf{$DF$} \\ \hline \hline
	1&the&116448129&the&2662742 \\ \hline
	2&of&59869301&and&2635683 \\ \hline
	3&and&58521777&to&2620998 \\ \hline
	4&to&53923142&of&2613789 \\ \hline
	5&a&40940103&a&2582936 \\ \hline
	6&in&36463498&in&2533431 \\ \hline
	7&is&22389310&for&2427073 \\ \hline
	8&for&21754176&is&2321630 \\ \hline
	9&that&16665399&on&2261602 \\ \hline
	10&on&15636014&with&2221913 \\ \hline
	11&with&13985141&are&1996578 \\ \hline
	12&it&13518855&this&1981575 \\ \hline
	13&be&13007008&from&1964174 \\ \hline
	14&as&11943257&be&1951862 \\ \hline
	15&are&11571176&by&1947784 \\ \hline
	16&you&11298405&as&1947590 \\ \hline
	17&this&11218852&that&1943238 \\ \hline
	18&by&10639772&at&1927033 \\ \hline
	19&at&9907466&it&1857296 \\ \hline
	20&i&9855628&an&1788905 \\ \hline
  \end{tabular}
 \label{tab:tc_df_20}
\end{table}
Since Table \ref{tab:tc_df_20} mainly holds stop words we show terms ranked between $101$ and $120$ and their $TC$ and $DF$ values
in Table \ref{tab:tc_df_120}. The correlation is obviously less strong and the number of intersecting terms went up to $33$.
\begin{table}
 \centering
  \caption{Terms ranked between $101$ and $120$ and their $TC$ and $DF$ Values}
  \begin{tabular}{|c||c|c||c|c|} \hline
        \textbf{Rank}&\textbf{Term}&\textbf{$TC$}&\textbf{Term}&\textbf{$DF$} \\ \hline \hline
	101&...&1667105&he&695825 \\ \hline
	102&get&1639959&get&689949 \\ \hline
	103&good&1623519&part&686865 \\ \hline
	104&her&1594657&need&684872 \\ \hline
	105&me&1578177&his&683221 \\ \hline
	106&back&1547902&could&680212 \\ \hline
	107&uk&1538433&those&678384 \\ \hline
	108&made&1524567&before&671783 \\ \hline
	109&way&1498196&between&671402 \\ \hline
	110&need&1498142&here&667614 \\ \hline
	111&those&1489151&available&660078 \\ \hline
	112&between&1484492&each&659960 \\ \hline
	113&she&1482487&n't&655339 \\ \hline
	114&2&1465266&back&644124 \\ \hline
	115&1&1462262&much&638500 \\ \hline
	116&day&1451911&used&634636 \\ \hline
	117&service&1439068&including&633138 \\ \hline
	118&world&1437539&help&632119 \\ \hline
	119&here&1436429&number&616825 \\ \hline
	120&used&1429151&own&614981 \\ \hline
  \end{tabular}
 \label{tab:tc_df_120}
\end{table}
\section{Experiment Results} \label{sec:exp_results}
\subsection{Correlation within the WaC Corpus}
\begin{figure}
 \centering
 \epsfig{file=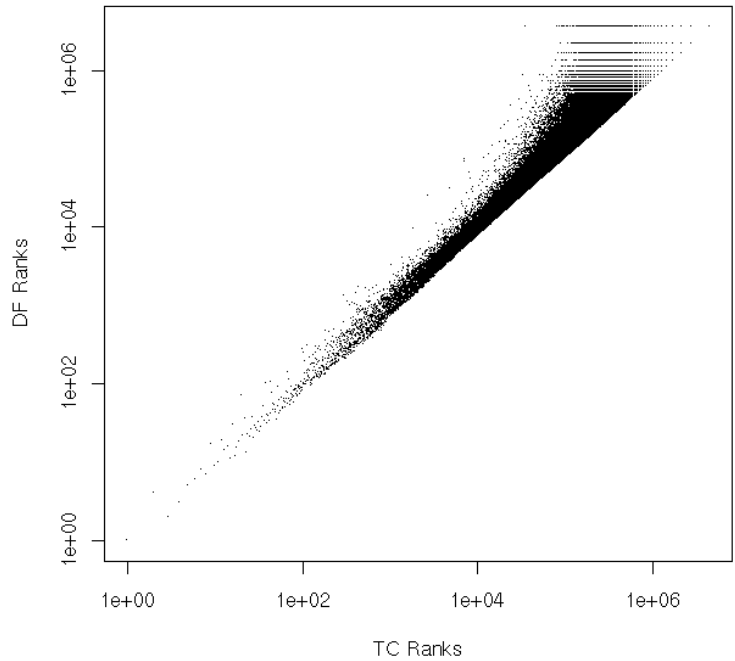,scale=0.5}
 \caption{Correlation between Term Count and Document Frequency in the WaC dataset}
 \label{fig:wac_ties_loglog}
\end{figure}

Figure \ref{fig:wac_ties_loglog} shows (in loglog scale) the correlation of ranked terms within the WaC corpus.
The x-axis represents the $TC$ ranks of terms and the y-axis the corresponding $DF$ rank of the same term.
As expected we see the majority of the points within a diagonal corridor which indicates a great similarity between the rankings.

\begin{figure}
 \centering
 \epsfig{file=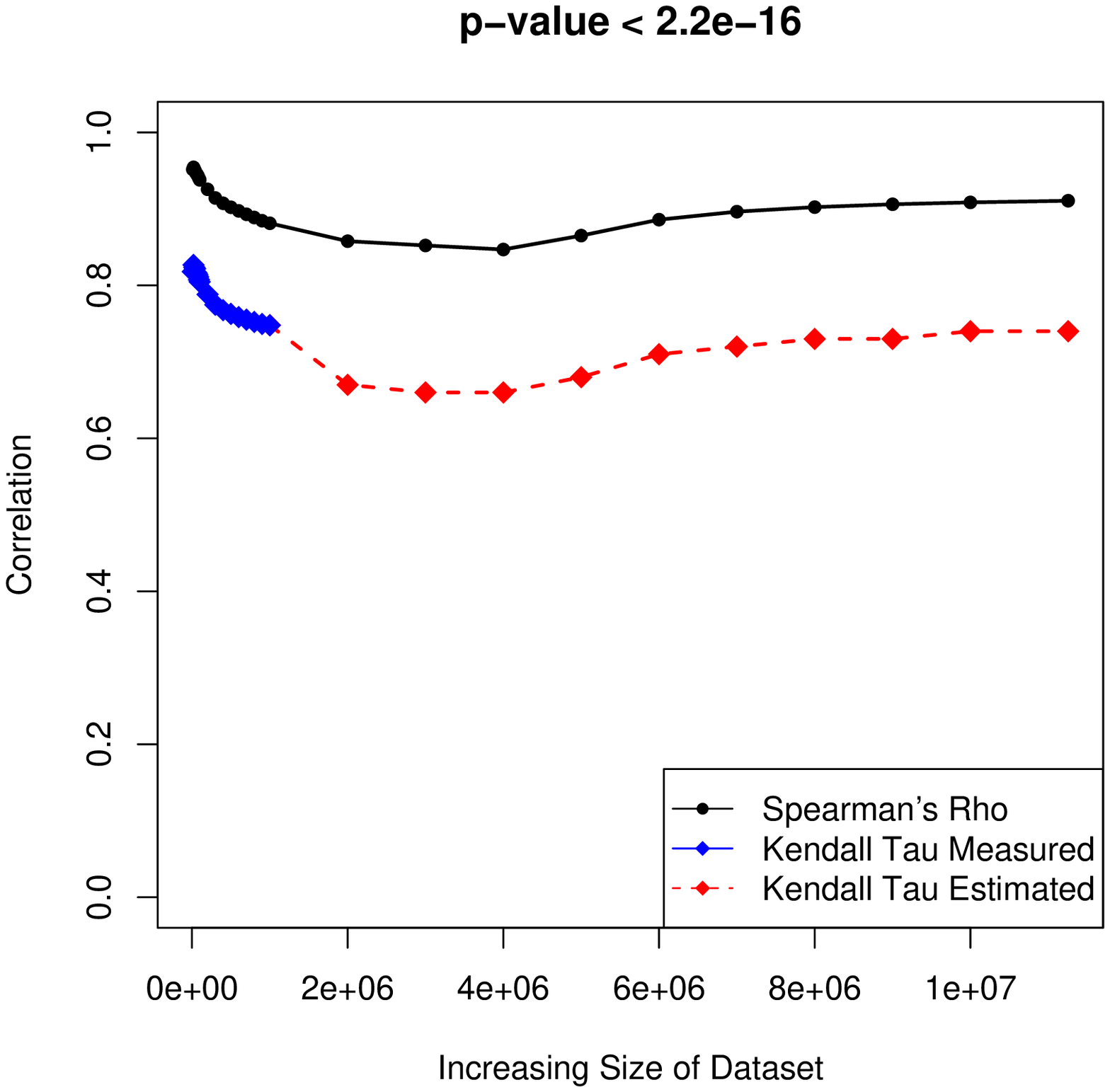,scale=0.45}
 \caption{Measured and Estimated Correlation between Term Count and Document Frequency in the WaC dataset - Normal Scale}
 \label{fig:spear_kend_corr_norm}
\end{figure}

\begin{figure}
 \centering
 \epsfig{file=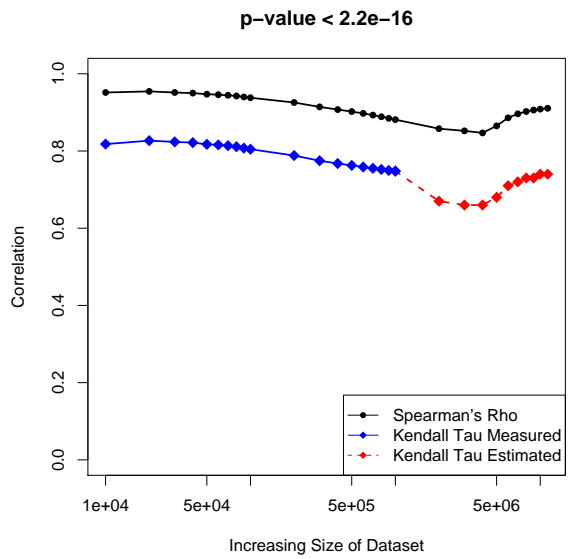,scale=0.45}
 \caption{Measured and Estimated Correlation between Term Count and Document Frequency in the WaC dataset - Semi-Log Scale}
 \label{fig:spear_kend_corr_log}
\end{figure}

Figures \ref{fig:spear_kend_corr_norm} and \ref{fig:spear_kend_corr_log} show the measured and estimated correlation between $TC$ and
$DF$ values in the WaC dataset.
The solid black line displays Spearman's $\rho$.
The increasing size of the dataset is shown on the x-axis.
The value for $\rho$ at any size of the dataset is beyond $0.8$ which indicates a very strong correlation between the rankings.
The results are statistically significant with a p-value of $2.2\times10^{-16}$.
%
%
The blue line in both Figures shows the computed Kendall $\tau$ values for the top $1,000,000$
ranks and the dotted red line represents the estimated values for the remaining set of data in the WaC corpus. Since the 
computed $\tau$ values are hard to read on a normal scale (Figure \ref{fig:spear_kend_corr_norm}) we plotted the same graph in semi-log
scale in Figure \ref{fig:spear_kend_corr_log}.
The computed $\tau$ values vary between $0.82$ and $0.74$ and the estimated values have a minimum of $0.66$.

We did not compute $\tau$ for greater ranks since it is a very time consuming operation and the estimated values
also indicate a strong correlation. 
Gilpin \cite{gilpin_kend_spear_table} provides a table for converting $\tau$ into $\rho$ values. We use this data to estimate our $\tau$
values. 
Even though the data in \cite{gilpin_kend_spear_table} is based on $\tau$ values computed from a dataset with bivariate normal population
(which we do not believe to have in the WaC dataset), it supports our measured values. 
For example, it shows that a $\tau$ value of $0.8$ can be converted to a $\rho$ of $0.94$ which is consistent with our measured values shown in
Figure \ref{fig:spear_kend_corr_norm}.
Therefore we can predict the high $\tau$ values even beyond the top $1,000,000$ ranks shown in Figure \ref{fig:spear_kend_corr_log}.
\subsection{Computation Time for Kendall $\tau$}
\begin{figure}
 \centering
 \epsfig{file=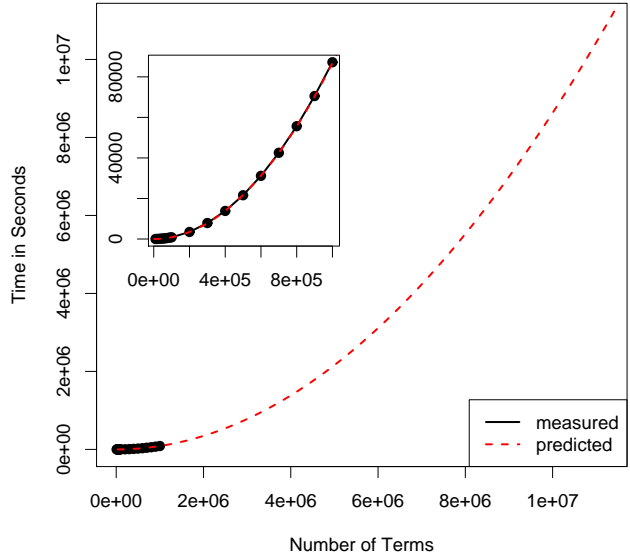,scale=0.5}
 \caption{Computation Time for Kendall Tau}
 \label{fig:kendall_time}
\end{figure}

Figure \ref{fig:kendall_time} shows the measured and predicted computation time (y-axis, in seconds) for $\tau$ of top $n$ rankings (x-axis).
The black solid line shows the measured time values for rankings up to the top $1,000,000$ terms. 
The red dashed line represents the predicted time values for the entire corpus and (in the small plot in the left top corner) for the 
top $1,000,000$ ranks. Figure \ref{fig:kendall_time} shows the observed complexity of $O(n^2)$.
For the entire WaC dataset (over 11 million unique terms) we estimate a computation time for Kendall $\tau$ of almost $11$ million seconds
or more than $126$ days which is clearly beyond a reasonable computation time for a correlation value.
Kendall $\tau$ was computed using an off-the-shelf correlation function as part of the \textit{R-Project}\footnote{\texttt{http://www.r-project.org/}},
an open source environment for statistical computing. The software (version 2.6) was run on a Dell Server with a Pentium P4 $2.8$Ghz CPU and
$1$ GB of memory.
\subsection{Term Count - Document Frequency Ratio in the WaC Corpus}
\begin{figure}
 \centering
 \subfigure[Rounded to two Decimals]{\label{fig:ratio_02}\epsfig{file=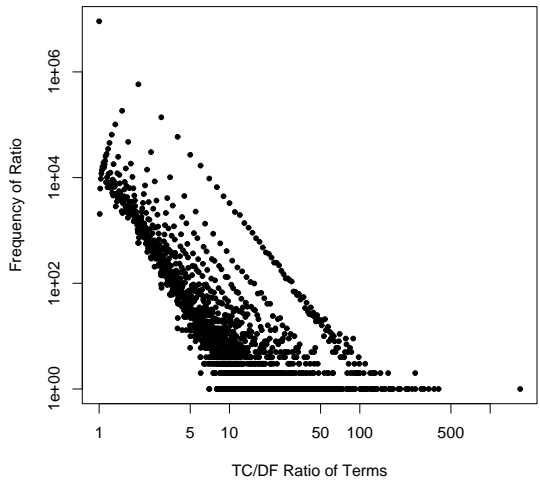,totalheight=3in}}
 \subfigure[Rounded to one Decimal]{\label{fig:ratio_01}\epsfig{file=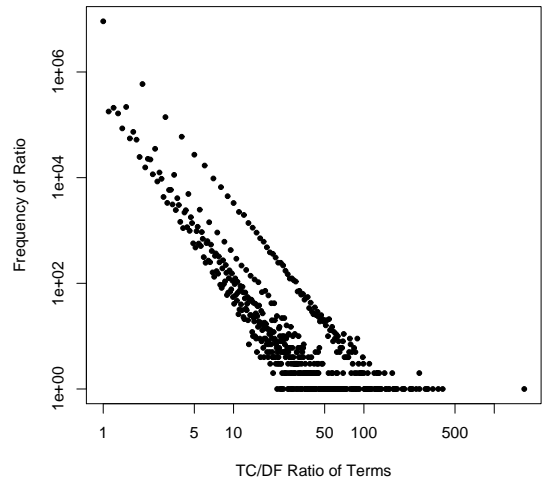,totalheight=3in}}
 \subfigure[Rounded to Integer Values]{\label{fig:ratio_int}\epsfig{file=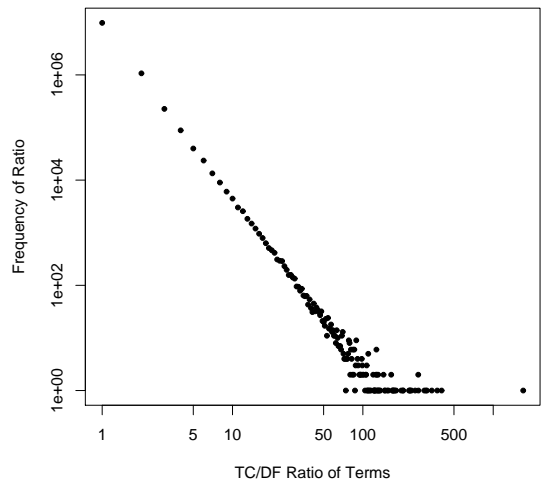,totalheight=3in}}
 \caption{Frequency of $TC$/$DF$ Ratios in the WaC Corpus}
 \label{fig:ratio}
\end{figure}

Another interesting way to show the correlation between $TC$ and $DF$ values is simply looking at the ratio of the two values.
Figure \ref{fig:ratio_02} shows the distribution of $TC$/$DF$ ratios with values rounded after the second decimal
and Figure \ref{fig:ratio_01} shows the ratios rounded after the first decimal. It becomes obvious that the
vast majority of the ratio values are very small. The visual impression is supported by the computed mean value 
of $1.23$ with a standard deviation of $\sigma=1.21$ for both, Figure \ref{fig:ratio_02} and \ref{fig:ratio_01}. The median of ratios is $1.00$ and $1.0$
respectively.
Figure \ref{fig:ratio_int} shows the distribution of $TC$/$DF$ ratios rounded as integer values. It is consistent with the pattern of Figures \ref{fig:ratio_02}
and \ref{fig:ratio_01} and the mean value is equally low at $1.23$ ($\sigma=1.22$). The median here is also $1$.
Figure \ref{fig:ratio} together with the computed mean and median values accounts for another solid indicator for the strong correlation between $TC$ and $DF$
values within the corpus.
\subsection{Correlation between the WaC and the N-gram Corpus}
\begin{figure}
 \centering
 \epsfig{file=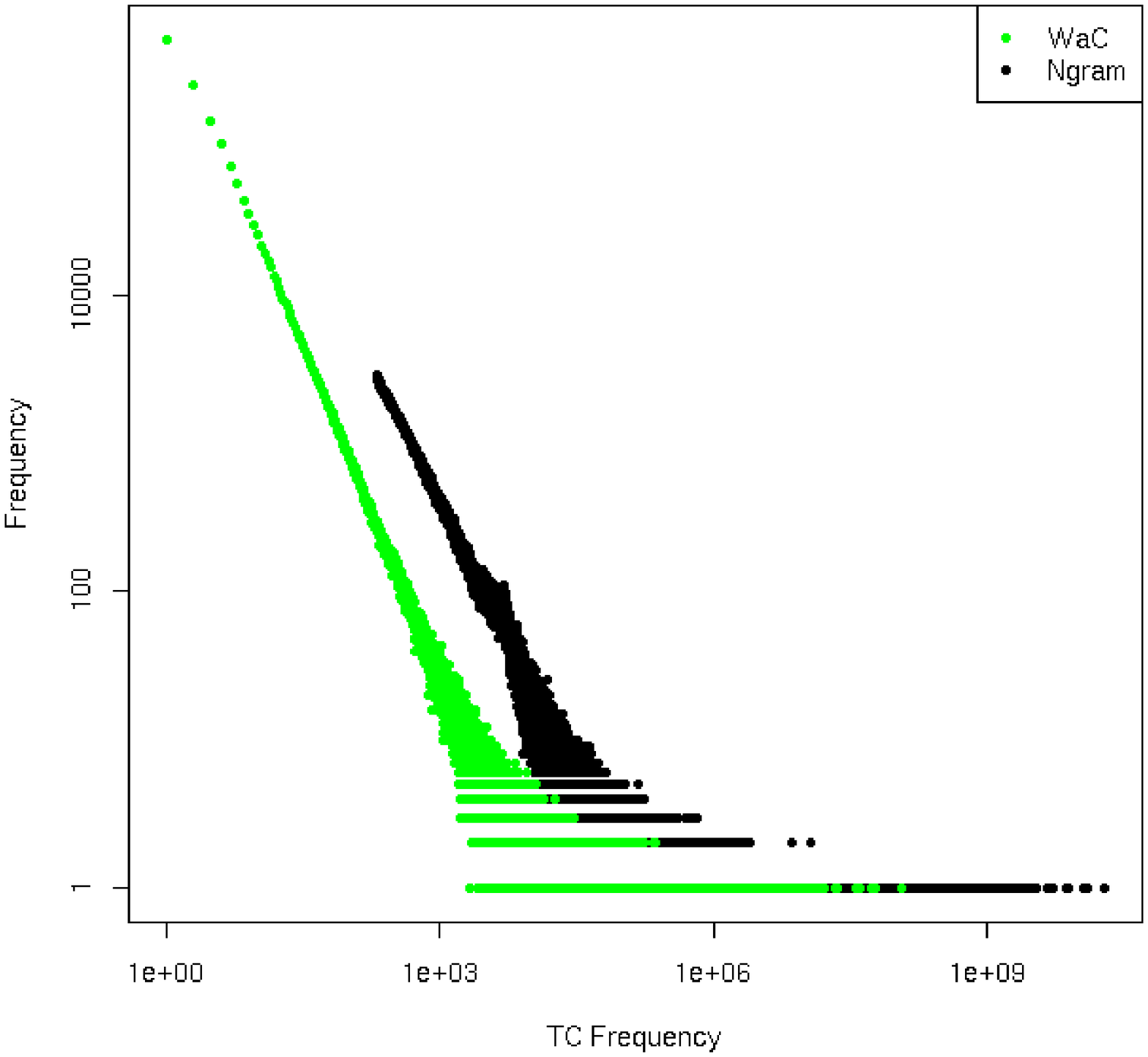,scale=0.6}
 \caption{Term Count Frequencies in the WaC and N-gram Corpus}
 \label{fig:tc_freqs}
\end{figure}

The $TC$ values for both corpora, WaC and N-gram, are available and therefore we investigate their correlation.
Figure \ref{fig:tc_freqs} displays (in loglog scale) the frequencies of unique $TC$ values in both corpora.
The graph shows the $TC$ threshold of $200$ Google applied while creating the N-gram.
By visual observation it becomes obvious that the distribution of $TC$ values in both corpora is very similar.
Just the size of the Google N-gram corpus is responsible for the offset between the graphs.
%
\section{Conclusion}
We have shown a very strong correlation between the $TC$ and $DF$ values within the WaC corpus with Spearman's $\rho\ge0.8$ ($p\le2.2\times10^{-16}$).
This result leads us to the conclusion that the two values can be used interchangeably and therefore $TC$ values are usable for the generation of
accurate IDF values.
We also show (by visual observation) a high correlation between the $TC$ values of the WaC and of the N-gram datasets.
We can now claim that, despite the fact that the Google N-gram dataset does not contain $DF$ values, the corpus and its $TC$ values are also usable
for accurate IDF computation which can lead to the generation of LSs of web pages.
\section{Acknowledgements}
We thank the Linguistic Data Consortium, University of Pennsylvania and Google, Inc. for providing the ``Web 1T 5-gram Version 1'' dataset.
We also thank the WaCky community for providing the ukWaC dataset.
Further we would like to thank Thorsten Brants from Google Inc. for promptly answering our emails and helping to clarify questions on the
Google N-gram corpus.
\newpage
%
%

%
\end{document}